# Tenets of QuantCrit

Ben Van Dusen[1] & Jayson Nissen[2]

[1]School of Education, Iowa State University, Ames, Iowa 50011, USA
[2]Nissen Education Research and Design, Corvallis, Oregon 97333, USA

Quantitative Critical (QuantCrit) research is a relatively new field of study grounded in critical theory (Crenshaw, 1990; Ladson-Billings, 2006; 2013). The nascency of QuantCrit has led multiple scholars to propose various related tenets (e.g., Gillborn *et al.*, 2018; Stage, 2007; Covarrubias *et al.*, 2018; López *et al.*, 2018). In this paper, we offer a synthesis of several tenets of QuantCrit we have applied to our research, their applications to education research in general, and citations for more information.

1. **The centrality of oppression**

*Definition*: QuantCrit researchers take as a fact that structural racism and sexism plague the U.S. economic, political, and educational systems. Gaps in student performance result from systems-wide policies and approaches that implicitly and explicitly disadvantage broad groups of students, particularly those that do not identify as white men. Students are oppressed for a myriad of dimensions of diversity beyond gender and race, including ableness, religious affiliation, and socioeconomic status. QuantCrit researchers commit to disrupting narratives that frame minoritized students as deficient and disrupting oppressive systems through anti-racist and anti-sexist work.

*Example research application*: When discussing the causes of inequities, QuantCrit researchers don't have to speculate about the causes. By a priori stating that the causes are racist, sexist, and classist power structures, researchers can focus their discussion on identifying the mechanisms and impacts of these oppressive systems.

2. **Data and methods are not neutral**

*Definition:* Researchers often fail to adequately discuss the biases introduced in data collection and analysis methods. This leads many researchers and readers to interpret quantitative findings as objective facts. QuantCrit research acknowledges that all data and analysis methods introduce biases and strives to minimize and explicitly discuss these biases.

*Example research application:* QuantCrit researchers avoid biasing findings by critically examining commonly used methods. For example, using *p*-values as go-no-go tests to identify differences between groups is particularly problematic in equity research. *P*-values depend on sample size. As many minoritized groups are underrepresented in the sciences, collecting sufficiently large sample sizes to find statistically significant differences between students who are minoritized and white men is prohibitive for many research projects. The lack of data from minoritized students has led many research projects to find that racism's impacts are not statistically significant and conclude that equity was achieved. Instead of using *p*-values to represent uncertainty, QuantCrit researchers use confidence intervals to create a more nuanced and less biased interpretation of findings.

   3. **Data cannot speak for itself**

*Definition:* When researchers present data or findings without explicitly providing a perspective, readers will likely interpret the data and findings through the dominant perspective, which often leads to racist and sexist interpretations (see Tenet 1). Such interpretations reinforce existing deficit narratives about minoritized groups.

*Example research application:* When discussing differences between demographic groups, QuantCrit researchers don't refer to them as the impacts of gender or race gaps but the impacts of sexism and racism. One way to do this in educational settings is to frame the differences as educational debts (Ladson-Billings, 2006) that society owes the students.

   4. **Groups are neither natural nor inherent**

*Definition:* In U.S. society, we often take one's racial and gender identity as being immutable features of a person. QuantCrit research reframes these aspects of individuals' identities as being socially constructed and fluid. Statistical analysis requires aggregation of data to support claims about group outcomes. QuantCrit research strives to create models that respect students' identities.

*Example research application:* QuantCrit researchers minimize student data aggregation, representing as much diversity in student outcomes as their data can reasonably allow. When aggregating data, QuantCrit researchers should do so in transparent ways that do not erase students and respect their identities.

   5. **Taking an intersectional perspective**

*Definition:* Identity is multifaceted (e.g., race, gender), each aspect of which exists along an axis. The interaction at the intersection of these axes shapes an individual's experience of the world. For example, Black women experience racism differently from Black men and sexism differently from White women.

*Example research application:* When developing statistical models, QuantCrit researchers explore the additional information that including interaction terms for demographic variables provides. By having an interaction term for gender and race, a model can predict the impacts of sexism and racism in ways that are not merely additive.

### 6. Valuing narrative and counter-narrative

*Definition:* Critical theory places great value on individuals' lived experiences. Narratives and counter-narratives capture these experiences. Counter-narratives represent the experiences and perspectives of minoritized people and often contradict our culture's dominant narratives (e.g., our educational system is meritocratic). QuantCrit researchers strive to include people's voices from minoritized groups in the data and research teams to ensure diversity in the narratives promoted.

*Example research application:* QuantCrit researchers can incorporate counter-narrative in a variety of ways. For example, QuantCrit researchers can integrate testimonios into a mixed-methods analysis (Covarrubias et al., 2018). QuantCrit researchers also strive to develop research teams with diverse experiences, perspectives, and identities. In projects where the research team lacks diversity, hiring an equity consultant to perform an audit of the work can provide a needed perspective.

**Final note:** There are other tenets of critical theory that inform our work, but we do not include those here for the sake of brevity (e.g., interest convergence). In creating this list, we focused on the tenets that have been most influential in shaping our QuantCrit research. More information can be found at stemequity.net.

**Acknowledgements**
This document is a product of NSF award # 1928596.**References**
Covarrubias, A., Nava, P. E., Lara, A., Burciaga, R., Vélez, V. N., & Solorzano, D. G. (2018). Critical race quantitative intersections: A testimonio analysis. Race Ethnicity and Education, 21(2), 253-273.
Crenshaw, K. (1990). Mapping the margins: Intersectionality, identity politics, and violence against women of color. Stan. L. Rev., 43, 1241.
Gillborn, D., Warmington, P., & Demack, S. (2018). QuantCrit: education, policy, 'Big Data' and principles for a critical race theory of statistics. Race Ethnicity and Education, 21(2), 158-179.
Ladson-Billings, G. (2006). From the achievement gap to the education debt: Understanding achievement in US schools. Educational researcher, 35(7), 3-12.
Ladson-Billings, G. (2013). Critical race theory—What it is not!. In Handbook of critical race theory in education (pp. 54-67). Routledge.3

López, N., Erwin, C., Binder, M., & Chavez, M. J. (2018). Making the invisible visible: advancing quantitative methods in higher education using critical race theory and intersectionality. Race Ethnicity and Education, 21(2), 180-207.

Stage, F. K. (2007). Answering critical questions using quantitative data. New directions for institutional research, 2007(133), 5-16.